\documentclass[eqsecnum,twocolumn,tightenlines,floats,floatfix,prc,nofootinbib,preprintnumbers]{revtex4}

\def\bea{\begin{eqnarray}}
\def\eea{\end{eqnarray}}

\def\st#1{{\kern-4pt} \not\!#1}

\def\sp{\kern +3pt}
\def\sm{\kern -3pt}
\def\spQ{\kern +6pt}

\usepackage[usenames]{color}

\def\be{\begin{equation}}
\def\ee{\end{equation}}
\def\ba{\begin{eqnarray}}
\def\ea{\end{eqnarray}}

\usepackage{graphics}
\usepackage{graphicx}
\usepackage{epsf}
\usepackage{amsmath}
\usepackage{amssymb}
\usepackage{slashed}
\usepackage{bm}

\def\sfrac#1#2{{\textstyle \frac{#1}{#2}}}

\begin{document}

\phantom{0}
\vspace{-0.2in}
\hspace{5.5in}


\preprint{{\bf ITEP-TH-08/17}} 
\vspace{.4cm}
\preprint{{\bf LFTC-18-5/26}}

\vspace{-1in}

\title
{\bf
Valence quark contributions for the
$\gamma^\ast N \to N(1440)$ form factors \\
from Light-Front holography}
\author{G.~Ramalho$\,^{1,2}$ and D.~Melnikov$\,^{1,3}$
\vspace{-0.1in} }

\affiliation{$^1$International Institute of Physics,
Federal University of Rio Grande do Norte,
Campus Universit\'ario - Lagoa Nova,  CP.~1613
Natal Rio Grande do Norte 59078-970, Brazil
\vspace{-0.15in}}

\affiliation{$^2$Laborat\'orio de F\'{i}sica Te\'orica e Computacional -- LFTC,
Universidade Cruzeiro do Sul, 01506-000, S\~ao Paulo, SP, Brazil
\vspace{-0.15in}}

\affiliation{$^3$Institute for Theoretical and Experimental Physics,
B.~Cheremushkinskaya 25, 117218 Moscow, Russia}

\vspace{0.2in}
\date{\today}

\phantom{0}

\begin{abstract}
The structure of the nucleon and the first radial excitation of
the nucleon, the Roper, $N(1440)$,
is studied  within the formalism
of Light-Front holography.
The nucleon elastic form factors and
$\gamma^\ast N \to N(1440)$ transition form factors
are calculated under the assumption
of the dominance of the valence quark degrees of freedom.
Contrary to the previous studies,
the bare parameters of the model associated
with the valence quark are
fixed by the empirical data for
large momentum transfer ($Q^2$) assuming
that the corrections to the three-quark picture
(meson cloud contributions) are suppressed.
The $\gamma^\ast N \to N(1440)$ transition form factors
are then calculated without any adjustable parameters.
Our estimates are compared with results from models based
on valence quarks and others.
The model compares well with the
$\gamma^\ast N \to N(1440)$ transition form factor data,
suggesting that meson cloud effects are not large,
except in the region $Q^2< 1.5$ GeV$^2$.
In particular, the meson cloud contributions
for the Pauli form factor are small.
\end{abstract}

\vspace*{0.9in}  
\maketitle

\section{Introduction}

Within the nucleon excited states ($N^\ast$)
the $N(1440)$ resonance,
also known as Roper, plays a special role.
Contrarily to the $\Delta(1232)$ and other
nucleon excitations, the Roper was not
identified as a bump in a reaction cross-section
but was instead found in the analysis of
phase-shifts~\cite{Roper64}.
Nowadays there is evidence that the Roper
should be identified as the first radial excitation
of the nucleon quark core,
although meson excitations are also
important for the internal structure.

Calculations based on valence quark degrees of freedom
are consistent with the
$\gamma^\ast N \to N(1440)$ transition form factors
for large $Q^2$ ($Q^2 > 2$
GeV$^2$)~\cite{Aznauryan07,Roper,Roper2,Roper-APS,Segovia15,Santopinto12}.
However, estimates based exclusively
on quark degrees of freedom
fail to describe the small $Q^2$ data
($Q^2 < 2$ GeV$^2$)~\cite{Aznauryan07,NSTAR,Aznauryan12}.
The gap between valence quark models and the data
at low $Q^2$ has been interpreted
as the manifestation of the
meson cloud effects~\cite{Roper,Roper2,Segovia15,Aznauryan12b}.
When the meson cloud contributions
are included in quark models
the estimates approach
the data~\cite{Cano98,Tiator04,Li06,Chen08,Golli09,Obukhovsky11}.

Besides, the calculations based on
dynamical coupled-channel reaction models,
where the baryon excitations are
described as baryon-meson states
with extended baryon cores~\cite{Burkert04,EBAC},
corroborates the importance of the meson cloud effects.
In those  models the mass associated
with the Roper bare core is about 1.7 GeV.
Only when the meson cloud dressing is considered
the Roper mass is reduced to the experimental value~\cite{Suzuki10}.
The interpretation of the Roper
as a radial excitation of the nucleon
combined with
a dynamical meson cloud dressing
solves the long standing problem
of the Roper mass in the context of a quark model~\cite{Capstick00}.

The Roper decay widths into $\gamma N$, $\pi N$ and $\pi \pi N$
are large comparative to other $N^\ast$ decays.
Those decay widths are also difficult
to explain in the context of a quark model.
The meson cloud dressing helps to explain
the $N(1440) \to \gamma N$ width,
where the contributions associated
with baryon-meson-meson states
play an important role~\cite{Suzuki10,Gegelia16,Golli09,Chen08}.

Overall the recent developments in the study
of the Roper electromagnetic structure
point to the picture of
a radial excitation of the nucleon
surrounded by a cloud of
mesons~\cite{Suzuki10,Golli09,Bauer14,Wang17,Roberts14a,Aznauryan12b}.
There is therefore a strong motivation to
study the Roper internal structure and
to disentangle the effects of the valence quark component
from the meson cloud component.
In particular, one can use the
knowledge of the baryon core structure
to infer the contribution due to the meson cloud.
This procedure was used in previous works
based on different frameworks
for the baryon core~\cite{Roper,Roper2,Segovia15,Aznauryan12b}.

In the case of the nucleon the
valence quark degrees of freedom
produce the dominant effect in the elastic form factors.
The meson cloud contribution to the nucleon
wave function is estimated to be
of the order of a few percent~\cite{Perdrisat07,Nucleon,Octet,Medium}.
As for the Roper, the meson cloud seems to play a more prominent role.
We conclude at the end that
the valence quark degrees of freedom
provide a very good description of the data,
but the meson cloud contribution are important
below 1.5 GeV$^2$, particularly for the Dirac form factor.

In the present work we propose
a new framework to analyze the role
of the valence quarks and the meson cloud in the Roper.
We use Light-Front holography
to estimate the leading order
(lowest Fock state)  contribution
to the $\gamma^\ast N \to N(1440)$ transition form factors.
Since the leading order calculation
includes only pure valence quark effects
in the baryon wave functions
($qqq$ contributions),
there is no contribution from the meson cloud.
In those conditions the gap between
the calculations and the data must
be essentially the consequence of the meson cloud effects.

The Light-Front (LF) formalism is
particularly appropriate to study
hadron systems, ruled by QCD,
and to describe the hadronic structure in
terms of the constituents~\cite{Teramond09a,Brodsky15,Teramond12a,Teramond11a}.
The LF wave functions (LFWF)
are relativistic and frame independent~\cite{Brodsky98,Brodsky15}.
The connection between LF quantization
of QCD and anti-de Sitter conformal field theory (AdS/CFT)
leads to Light-Front holography~\cite{Teramond12a,Brodsky15,Brodsky08a}.
LF holography have been used
to study the structure of hadron properties,
such as the hadron mass spectrum,
parton distribution functions, meson and
baryon form factors
etc.~\cite{NSTAR,Teramond12a,Chakrabarti14a,Brodsky15,Teramond13a,Teramond15a,Forkel07,Karch06,Ballon12,Grigoryan09,Grigoryan07a,Bayona15a,Brodsky06,Brodsky08a,Branz10a,Vega11,Ahmady16,Mamedov16,NucleonAxial,Vega11a}.
In particular, the  formalism was recently applied to the study of the
nucleon~\cite{Hong08a,Abidin09,Gutsche12a,Chakrabarti13a,Liu15a,Maji16a,Sufian17a} and Roper~\cite{Teramond11a,Teramond12a,Gutsche13a}
electromagnetic structure.

An important advantage of the LF formalism
applied to hadronic physics
is the systematic expansion of the wave functions
into Fock states with
different number of constituents~\cite{Brodsky01,Brodsky98,Brodsky15}.
In the case of the baryons the leading order
contributions is restricted to the three-valence quark configuration.
Although the restriction to the lowest order Fock state
(three valence-quark system) may look
as a rough simulation of the real world,
it may provide an excellent first approximation
when the confinement is included in the LFWF,
defined at the Light-Front time~\cite{Brodsky08a,Perdrisat07,Teramond15a}.
In those conditions non-perturbative physics is
effectively taken into account
by LF holography~\cite{Brodsky15,Teramond15a}.

Under the assumption that LF holography
can describe accurately the large-$Q^2$ region
dominated by valence quark degrees of freedom,
we calibrate the free parameters of the model
by the available data above a given threshold $Q^2 \ge Q_m^2$
($Q_m^2 = 1.5$--2.5 GeV$^2$).
This procedure differs from the previous
studies where the free parameters,
associated with the nucleon anomalous
magnetic moments were fixed at $Q^2=0$,
where the meson cloud contamination is
expected to be stronger.
Once the free parameters are fixed by the
nucleon data, one uses the model to calculate the
$\gamma^\ast N \to N(1440)$ transition form factors.
Since no parameter is adjusted by the
Roper data, our calculations are true predictions
for the $\gamma^\ast N \to N(1440)$ transition form factors.

This article is organized as follows:
In Sec.~\ref{secHolography} we review the LF setup originally used in Ref.~\cite{Abidin09}
in the context of calculation of baryon form factors.
In Secs.~\ref{secNucleon} and~\ref{secRoper} we adapt
the calculations of Refs.~\cite{Gutsche12a,Gutsche13a} to derive our main results.
In particular, in Sec.~\ref{secNucleon},
we review the results for the nucleon form
factors and fix the parameters by the nucleon data.
{Sec.~\ref{secRoper} contains our}
results for the valence quark contribution
to the $\gamma^\ast N \to N(1440)$ transition form
factors and their discussion.
The outlook and conclusions are presented in Sec.~\ref{secConclusions}.

\section{Light-Front holography}
\label{secHolography}

It is by now well known that string theory or gravity in anti-de Sitter
(AdS) space can provide a description of some lower-dimensional
(conformal) gauge theories (CFT)
at strong coupling~\cite{Maldacena97,Gubser98,Witten98}.
This AdS/CFT correspondence, or holography, can,
with certain important restrictions, be applied to QCD-like theories,
e.g.~\cite{Witten98b,Klebanov00,Sakai04}.
It remains an open question, however,
whether the \emph{top-down} string theory methods
are practical enough to model QCD itself.
Instead it proved efficient to apply \emph{bottom-up}
approaches to QCD~\cite{Erlich05,DaRold05,Karch06}.
Such approaches are motivated phenomenologically by the string models,
but lack a first-principle justification.
Among those, the approach of LF holography is based
on a comparison of the results of the LF quantization
of QCD~\cite{DrellYan,West}
with the predictions of holographic (AdS/CFT)
models~\cite{Teramond12a,Brodsky15,Gutsche12a,Chakrabarti13a,Liu15a,Hong07a}.

\subsection{Light-Front QCD vs AdS/CFT}

Light-Front quantization provides a powerful tool
to study systems ruled by microscopic QCD dynamics.
It starts with introducing a special (Light-Front) parametrization of
the Hamiltonian, which consequently acts on
a Hilbert space of the LF wave functions spanned by partonic
Fock states associated with quarks and gluons~\cite{Teramond12a,Brodsky15}.

The Fock states $|n\rangle$ are multi-particle eigenstates of the
free Hamiltonian for the constituent quarks and gluons.
Their coefficients $\psi_n(\varkappa_i, {\bf k}_{\perp i})$
in the LFWF expansion depend on the fraction of momentum  $\varkappa_i$
of partons, $i=1,\ldots,n$, as well as the
partons transverse momenta ${\bf k}_{\perp i}$~\cite{Brodsky15,Brodsky06}.
The notation $\psi_n(\varkappa_i, {\bf k}_{\perp i})$ implies 
a dependence on the fractions $\varkappa_i$ and transverse momenta of 
all the partons.

In the standard terminology $|n\rangle$ are the states of a given
\emph{twist} $\tau\equiv n$.
The amplitudes $\psi_n(\varkappa_i,{\bf k}_{\perp i})$
define the probability to find the hadron in the given $n$-parton Fock state.
The Fock state expansion separates the dependence on bound state total
four-momentum $P$, contained in $|n\rangle$, from the
frame-invariant dependence on the relative variables
$\varkappa_i$ and ${\bf k}_{\perp i}$, which are subject to the conditions
$\sum_{i=1}^n \varkappa_i =1$ and $\sum_{i=1}^n{\bf k}_{\perp i}= {\bf 0}$~\cite{Liu15a,Brodsky15}.
A useful representation of amplitudes is in terms 
of the conjugate impact space, obtained by a 2D Fourier transform 
to $\widetilde  \psi_n(\varkappa_i, {\bf b}_{\perp i})$, which depends on 
the impact parameters ${\bf b}_{\perp i}$, satisfying 
$\sum_{i=1}^n{\bf b}_{\perp i}= {\bf 0}$~\cite{Brodsky06,Brodsky15}.

The relativistic LF eigenvalue equation $H_{LF} |\Psi_h \rangle = M_h^2 |\Psi_h \rangle $, for the LFWF $|\Psi_h\rangle$ of the hadron $h$, 
can be cast in a form of a Schr\"odinger-like equation
for  $\widetilde  \psi_n(\varkappa_i, {\bf b}_{\perp i})$, 
where the Hamiltonian $H_{LF}$ can be separated into particle kinematic terms 
and a complicated (unknown) interaction potential for the constituents. 
For practical purposes it is natural to reduce this 
hadron system down to a system where the parton 
is subject to an simplified one-body potential
which captures the effect of confinement~\cite{Liu15a,Brodsky15,Teramond09a}.

In the study of electromagnetic interaction 
in impulse approximation one can regard
a baryon as a system of an active quark and 
a cluster of $(n-1)$ spectator partons. 
In this conditions the amplitudes 
of the Fock states can be reduced to 
$\widetilde  \psi_n(\varkappa, {\bf b}_{\perp})$, where 
$\varkappa$ (respectively $1-\varkappa$) and 
${\bf b}_{\perp}$ ($-{\bf b}_{\perp}$) are the variables associated 
with the active quark (spectator cluster),
so that $\widetilde  \psi_n(\varkappa, {\bf b}_{\perp})$
can be interpreted as the wave function of a two-body
quark-cluster system~\cite{Liu15a,Sufian17a,Teramond09a,Brodsky15}.

In the case of the baryons, in leading twist approximations ($\tau =n =3$) 
and massless quarks, one can reduce the LF wave equation
to the form~\cite{Liu15a,Teramond09a,Brodsky15,Sufian17a}
\ba
\left[ - \frac{ \nabla^2_{ {\bf b}_{\perp} }} {\varkappa (1- \varkappa)} 
 + U_{\rm eff} 
\right] \tilde{\psi}_n( \varkappa, {\bf b}_{\perp} )  
= M_h^2 \tilde{\psi}_n( \varkappa, {\bf b}_{\perp}), 
\ea
where the effective potential $U_{\rm eff}=\mu_J + \Phi$ 
consists of the effective potential $\mu_J$, encoding the 
total angular momentum $J$ of the hadron, 
and the effective interaction potential $\Phi$, 
exhibiting confinement~\cite{Brodsky15}. 
The kinetic term in the equation also encodes the relative 
angular momentum $L$ of the 
quark-spectator system ($L=0,1$)~\cite{Liu15a,Brodsky06,Brodsky15}.

Furthermore, in the semiclassical approximation,
when the quarks have no mass and the quark loops are neglected, 
one can replace the dependence
on the variables $\varkappa$ and ${\bf b}_{\perp}$
in $\widetilde \psi_n$ by the dependence on a single parameter
$\zeta = \varkappa (1- \varkappa) |{\bf b}_{\perp}|$,
except for an overall factor
$f(\varkappa)=\sqrt{\varkappa(1 -\varkappa)}$~\cite{Brodsky06,Brodsky15,Liu15a}.
The LF wave equation associated with $\widetilde \psi_n(\varkappa, \zeta)$ 
then becomes a one-dimensional Schr\"odinger equation
in the ``radial'' variable $\zeta$.
The variable $\zeta$ measures the separation
between the active quark and the remaining spectator partons.
In the case of $n$-body systems $\zeta$ appears to be a $\varkappa$-weighted average of the impact variables: 
$|{\bf b}_{\perp}| \to |\sum_{j=1}^{n-1} \varkappa_j {\bf b}_{\perp j}|/|\sum_{j=1}^{n-1} 
\varkappa_j|$~\cite{Brodsky06,Brodsky15,Sufian17a}.

The key observation of LF holography is the equivalence of the Schr\"odinger 
equation for the LFWFs $\widetilde \psi_n(\varkappa, \zeta)$ and the characteristic 
equations of matter fields in five-dimensional anti-de Sitter (AdS$_5$) space,
assuming that one identifies $\zeta$ with
the radial coordinate of AdS$_5$, $z$~\cite{Brodsky06,Radyushkin95,Brodsky15}.
For that correspondence contributes the fact 
that the residual $\varkappa$ dependence in the wave function factorize,
and can be absorbed
in the normalization factors in the 5D theory calculation of matrix elements.
Consequently, the $\varkappa$ dependence is not relevant 
for the holographic correspondence.

The purpose of this work is to understand the leading twist
contribution ($\tau=3$) for the nucleon and Roper, 
which corresponds to the valence quark approximation to certain QCD processes.
To study  the electromagnetic properties of the nucleon radial excitations
we use the two-body decomposition of the three-quark systems ($\tau=3$),
representing those systems as an active quark and a spectator cluster 
with masses (eigenvalues) and wave functions determined 
by the holographic equivalents of LF wave equation,
as described above.  
The AdS wave equation 
used in the present work is discussed in the next sections.

\subsection{5D fermion in AdS space}

In the bottom-up holographic approach to QCD the nucleon and the nucleon excitations
can be introduced as fermion fields
in five-dimensional AdS space~\cite{Hong07a,Hong07}.
To define this curved space one requires the metric tensor $g$,
which can be conveniently introduced via
line element $ds^2$ in the Poincar\'e coordinates:
\be
\label{Eqmetric}
ds^2 \ = \ \frac{R^2}{z^2}\left(\eta_{\mu\nu}dx^\mu dx^\nu - dz^2\right),
\ee
where $\eta_{\mu\nu}=\text{diag}(+,-,-,-)$ is the 4D Minkowski
metric tensor and $R$ is a parameter called AdS radius.
The latter sets the scale
for the space's curvature. AdS$_5$ radial coordinate $z$
is to be identified with the LF parameter $\zeta$.

From the metric one can define the Dirac operator $\hat {\cal D}$,
which we write as
\ba
\hat {\cal D}
= \frac{i}{2}\,e^M_A\Gamma^A \left({\overleftrightarrow{\partial}_M} +
\sfrac{1}{8}
\omega_{M}^{AB}[\Gamma^A,\Gamma^B]\right).
\ea
Here $\Gamma^A$, $A,B=0,1,2,3,z$ is the standard set of five
Dirac gamma matrices and we choose to represent in the chiral basis.
Frame $e^M_A$ and spin-connection $\omega_M^{AB}$ tensor fields
can be computed from the metric tensor.
For completeness we summarize the explicit formulas here
\ba
e_{A}^M \Gamma^A & = & 
\frac{z}{R}(\gamma^\mu, - i \gamma_5)\,, \label{Eqframe}\\
\omega_M^{AB} & = & \frac{1}{z}
(\eta^{Az}\delta_M^B-\eta^{Bz}\delta_M^A) \,
\ea
It is also understood that
$A{\overleftrightarrow{\partial}}_M B = A (\partial_M B) - (\partial_M A) B.$

Substantiating the above claim, the kinematic properties of nucleons,
\emph{e.g.} the spectrum, can be described by a theory
of a massive fermion in AdS$_5$:
\ba
\label{Eqaction}
S_N &= &
\int d^4 x\, d z \ \sqrt{-\mbox{det} \; g} \;
\overline \Psi \big(\hat{{\cal D}}-\mu - \Phi \big) \Psi\,,
\ea
where $\mu$ is the 5D fermion mass to be fixed below.

The function $\Phi(z)$ is an effective scalar potential,
whose role is to introduce the complex dynamics of QCD,
essentially the confinement. When $\Phi=0$, the equations
of motion of the 5D theory can be cast in
the form corresponding to the LF Schr\"odinger equation
in the conformal limit~\cite{Brodsky15}.
A phenomenologically compelling choice
of the potential is~\cite{Abidin09,Gutsche12a}
\be
\label{potential}
\Phi=\kappa^2 z^2\,.
\ee
The dimensionful parameter $\kappa$ breaks conformal symmetry
and introduces a mass scale, which determines
the meson and baryon spectrum~\cite{Teramond12a,Brodsky15,Teramond15a}.
It can be thus related to $\Lambda_{\rm QCD}$.
A conventional way to introduce this potential
is through the dilaton field background~\cite{Abidin09,Gutsche12b}.
This procedure makes the connection with top-down
AdS/CFT constructions. The corresponding bottom-up case is
dubbed the soft-wall model.

In four dimensions the four-component Dirac spinor describes
two independent chirality modes. Although five-dimensional
Dirac spinor $\Psi$ equally has four components,
it will correspond to a nucleon with a specific single choice of chirality.
This is a consequence of boundary conditions necessary
in the holographic approach~\cite{Henningson98,Mueck98,Henneaux98,Abidin09,Gutsche12a,Hong07a}.
In order to preserve CP invariance, the second chirality mode
is introduced similarly as Eq.~(\ref{Eqaction}),
but with an opposite sign in front of $\mu$ and $\Phi$,
\emph{e.g.}~\cite{Gutsche12a}.
For simplicity of presentation we only review one chirality mode here.

\subsection{Wave equations and wave functions}

In order to solve the Dirac equations derived from action~(\ref{Eqaction})
the fermion field (spin 1/2 and positive parity) $\Psi$
is decomposed into its left- and right- chirality components:
\ba
\Psi(x,z) = \Psi_L(x,z) + \Psi_R(x,z),
\ea
where $\Psi_{L/R} = \sfrac{1}{2}(1 \mp \gamma^5)\Psi$.
The solution can then be found via separation of variables,
assuming a plane wave dependence on the 4D spacetime coordinates
\be
\Psi_{L/R}(x,z)\ = \ \psi_{L/R}(x)z^2F_{L/R}(z)\,,
\ee
where left and right Weyl spinors $\psi_{L/R}(x)=\psi_{L/R,P} \,e^{i P \cdot x}$
satisfy the 4D Dirac equation with mass $M^2=P^2$.
We also pull out the $z^2$ factor for convenience.
In the end, one is left with a pair of coupled equations
for scalar \emph{profile} functions $F_{L/R}(z)$:
\ba
\left[\pm\partial_z +
\frac{\mu R + \Phi}{z}\right] F_{L/R} (z) = M\,
F_{R/L}(z)\,.
\label{eqHolog1}
\ea

The equations must be solved with an appropriate choice
of boundary conditions. First, holography requires regularity
of the wave function $\Psi(x,z)$ as $z \to \infty$.
For $z\to 0$ there are two linearly independent solutions
$F_{L/R}\sim z^{\mp \mu R}$.
In holography one usually chooses to call one of those
solutions the source and another one the vev of the operator
dual to the bulk (5D)
field $\Psi$ -- in this case baryon interpolating operator.
For the problem in question it only makes sense to choose the
vev solution as $F_R \sim z^{\mu R}$ and the source $F_L=0$.
The opposite choice, source $F_R=0$ and vev
$F_L \sim z^{-\mu R}$ leads to a non-physical spectrum.
Note that for the opposite chirality case one changes $\mu\to - \mu$,
and the two choices of the $z\to 0$ boundary conditions
are interchanged, with the physical one being $F_R=0$ and
$F_L \sim z^{\mu R}$.

After fixing the boundary conditions it is standard to rewrite system
(\ref{eqHolog1}) as a second order Schr\"odinger-like
differential equation for either $F_R$, or $F_L$.
Since we are solving the boundary problem, any solution
can be expanded over an eigenfunction basis,
labeled by an integer $n$. Equation
\begin{multline}
-\frac{d^2}{d z^2}F_R + \frac{m(m - 1)}{z^2}
\,F_R + 2\kappa^2\left(m+\frac12\right)F_R
\\ +\kappa^4z^2F_R  =  M_n^2F_R,
\end{multline}
has a set of eigenfunctions $F_{L/R,n}(z)$ expressed in terms of
the generalized Laguerre polynomials $L_n^\alpha$:
\be
\label{eqFR}
F_{R,n}(z) \ \propto \ z^{m} e^{- \kappa^2z^2/2}L_n^{m-1/2}(\kappa^2z^2).
\ee
The second component $F_L$ can now be calculated using
Eqs.~(\ref{eqHolog1}):
\be
\label{eqFL}
F_{L,n}(z) \ \propto \ z^{m+1} e^{-\kappa^2z^2/2}L_n^{m+1/2}(\kappa^2z^2).
\ee
These eigenfunctions correspond to eigenvalues given by
\be
M_{n}^2= 4 \kappa^2 \left( n + m + \frac{1}{2}\right).
\label{Leigenvalues}
\ee
We recall that in the previous expressions $m\equiv \mu R $.

The eigenvalues provide the result for the 4D spectrum
of nucleon excitations with radial excitation quantum number $n$
and angular momentum quantum number linear in the parameter $m$,
which we will fix below.
The above choice of the potential~(\ref{potential}) leads
to the spectrum consistent with Regge
trajectories~\cite{Teramond12a,Brodsky15,Karch06}.

Summing up, the full solution for the 4D positive chirality mode reads
\be
\label{PosChirMod}
\Psi_n^+(x,z)=
z^2\left(\begin{array}{c}
F_{L,n}(z)\,\chi_n(x)   \\
F_{R,n}(z)\, \chi_n (x)  \\
\end{array} \right),
\ee
where $\chi_n(x)$ are two component spinors related
with the Weyl spinors:
$\psi_L^\dagger (x) = (\chi_n^\dagger (x), 0)$ and
$\psi_R^\dagger (x) = (0, \chi_n^\dagger (x))$.
The negative chirality modes is obtained
via an appropriate change of signs and exchange of the solutions for
$F_{L/R}$~\cite{Gutsche12a,Gutsche13a}:
\be
\label{NegChirMod}
\Psi_n^-(x,z)=
z^2\left(\begin{array}{c}
F_{R,n}(z)\,\chi_n(x)   \\
-F_{L,n}(z)\,\chi_n(x)  \\
\end{array} \right).
\ee

Using the component associated with the index $n$
we can write the fermion fields
with positive and negative parities as
\ba
\Psi^\pm (x,z) = \frac{1}{\sqrt{2}}
\sum_n \Psi_n^\pm (x,z).
\ea

\subsection{Interactions}

In hadron physics, information on the structure of nucleon
resonances is contained in the matrix elements of interaction currents between
the initial and final hadrons,
including the nucleon and the nucleon excitations (baryons).
In holography, 4D operators of conserved currents
$J_\mu$ are described by massless gauge fields $V_M$ in AdS$_{5}$ space.
Interacting bulk baryon fields $\Psi$ couple to the gauge fields,
which is a dual holographic description of the coupling of baryons
to conserved currents.
Matrix elements of the currents, can be computed from the interaction terms,
which are, schematically, overlaps of the bulk fields
with the source gauge field~\cite{Strassler01}:
\ba
\label{Sint}
S_{\rm int} = \int d^4 x\, d z\, \sqrt{-\det g} \;
\bar \Psi(x,z)\,
 \hat {\cal V}(x,z) \Psi (x,z).\quad
\ea
Here $\hat{\cal V}$ encodes a coupling of field $V_M$ to fermions.
We consider in particular the decomposition
\ba
 \hat {\cal V}(x,z) = \hat {\cal V}_0(x,z) +  \hat {\cal V}_1(x,z) +
 \hat {\cal V}_2(x,z),
\label{eqVtotal}
\ea
as discussed next.
For simplicity, we only present the interaction for positive chirality.
For the opposite chirality case one has to adjust the signs
in the above expression to preserve the CP invariance~\cite{Gutsche12a}.

The term $ \hat {\cal V}_0(x,z)$ is representing the minimal Dirac coupling
given by~\cite{Abidin09}
\ba
 \hat {\cal V}_0(x,z) = \hat Q \, \Gamma^M V_M (x,z),
\ea
where $V_M$ could be either the abelian dual of the electromagnetic current, or the non-abelian isovector current. Following the parametrization of~\cite{Gutsche12a,Gutsche13a} $\hat Q = e_N = \sfrac{1}{2}(1+ \tau_3)$,
the nucleon charge  ($N=p,n$), for the elastic electromagnetic transitions,
and $\hat Q= \tau_3$ for the transition between
the nucleon and higher mass $J^P = \frac{1}{2}^+$ resonances
($\tau_3$ is the Pauli isospin operator).
This \emph{minimal} coupling only yields the Dirac form factor.

To produce the Pauli form factor more input from the holographic side is necessary. Abidin and Carlson~\cite{Abidin09} proposed to consider a non-minimal extension of the coupling $\hat{\cal V}_0$, adding a term
\ba
\label{V1}
\hat{\cal V}_1 (x,z)\ =\  \frac{i}{4}\,\eta_N [\Gamma^M,\Gamma^N]V_{MN}(x,z)\,,
\label{eqV1}
\ea
where $V_{MN} = \partial_M V_N - \partial_N V_M + [V_M,V_N]$ and
one can consider different couplings $\eta_S$ and $\eta_V$
for the isoscalar and isovector parts respectively
[$\eta_N=\sfrac{1}{2}(\eta_S + \eta_V \tau_3)$].
As one may expect from its Lorentz structure, this term yields the Pauli
form factors, as well as a correction to the Dirac form factors.
Moreover, since the non-minimal term contains derivatives of
the field $V_M$ and extra powers of $z$, the Pauli form factor
turns out to be subleading with respect to the Dirac
form factor at large momentum transfer, as expected, while the correction
to the Dirac form factor is of the same order.
The effect of the term ${\cal V}_1$ on the nucleon form factors
is studied in
Refs.~\cite{Teramond12a,Chakrabarti13a,Abidin09,Gutsche12a,Gutsche13a,Maji16a}.
We provide additional details in what follows.

More couplings can be considered.
In Refs.~\cite{Gutsche12a,Gutsche13a} the following minimal-type
(isovector) coupling was suggested
\be
\label{V2}
\hat{\cal V}_2 (x,z) \ = g_V \tau_3 \, \Gamma^M\gamma^5V_M (x,z)\,.
\ee
where $g_V$ is a coupling constant.
This coupling is consistent with gauge invariance and discrete symmetries,
that can be added to the 5D action to improve
the fitting of experimental data~\cite{Gutsche12a}.

One first observation about this term is that it naively breaks CP in 4D.
However, this is not the case, since 5D fermion $\Psi$ describes
only one 4D chirality.
Adding a similar interaction term for the opposite chirality,
with the opposite sign of coupling, ensures 4D
CP invariance.

A more serious issue is that Eq.~(\ref{V2}) breaks 5D covariance. Adding such term requires an implicit background 5D vector field. Such background fields are more common in the higher-dimensional top-down holographic models with flux compactifications. Here we assume that such a flux compactification exists and we assess its effect on the observable form factors.
We further comment on the general structure of the term~(\ref{Sint})
in Ref.~\cite{inpreparation}.

In order to calculate transition matrix elements of the currents using Eq.~(\ref{Sint}) one needs to specify the 5D gauge field $V_M(x,z)$. Similarly to fermions, a bosonic gauge field dual to the isovector or isoscalar current is introduced by a 5D Lagrangian. Specifically, one needs to solve equations following from the action
\be
S_{J} \ = \ -\int d^4 x\, d z\, \sqrt{-\det g}\,\frac{e^{-\Phi}}{4}\,{\rm Tr\,}V_{MN}V^{MN}\,,
\ee
subject to appropriate boundary conditions.
These boundary conditions differ
from the boundary conditions for the fields
associated with the nucleon
and the nucleon excitations,
since they have to be of the \emph{source} type at $z=0$
(the leading solution as opposed to the subleading vev-type).
The regularity condition requires $V_M$ to vanish for $z\to \infty$.

In what follows we will be interested in
the abelian electromagnetic current, so the non-abelian details
of this discussion can be omitted.
Altogether, the solution can be cast
in the form~\cite{Grigoryan07b,Gutsche12a,Gutsche13a}
\ba
\label{Fourier}
V_\mu(x,z) \ = \ \int\frac{d^4q}{(2\pi)^4}\,e^{-i q \cdot x} \epsilon_\mu(q)V(-q^2,z)\,,
\ea
where $\epsilon_\mu$ is the photon polarization vector, and
\ba
\label{Vpropagator}
V(Q^2,z)= \kappa^2 z^2
\int_0^1 \frac{d \varkappa}{(1-\varkappa)^2}
x^{\frac{Q^2}{4 \kappa^2}} e^{- \frac{\kappa^2 z^2 \varkappa}{1-\varkappa}}\,,
\ea
with $Q^2=-q^2$.
In the limit $z \to 0$, one has $V(Q^2,0)=1$.

Substituting Eqs.~(\ref{Fourier}) and (\ref{Vpropagator}) together with
the modes~(\ref{PosChirMod}) and~(\ref{NegChirMod})
into the interaction Lagrangian~(\ref{Sint}) yields matrix elements
of the current up to some normalization terms, from which form factors
are read off directly.
We spare the reader the details of this analytical calculation,
but rather summarize the final results in Sec.~\ref{secNucleon} for the nucleon
and in Sec.~\ref{secRoper} for the Roper. See Refs.~\cite{Gutsche12a,Gutsche13a}
for the original derivation.

\subsection{Mass spectrum and more}

In this section we specify the details of the holographic model
and relating it to the observable spectrum of the nucleon radial
excitations and the $\rho$ mesons.

First, the nucleon ($\Psi_{N0}$) and the Roper ($\Psi_{N1}$)
are the states with radial quantum numbers $n=0,1$ and angular momentum $L=0$.
Any of these states must be regarded as a superposition of the given twist
(number of partons) Fock states, so we assume that
the 5D mass $\mu$ is encoding both the angular momentum $L$ and
the twist $\tau$, $\mu R = m(L,\tau)$.

A way to fix the relation for $L=0$ is to analyze the large $Q^2$ scaling of
$F_{L/R,n}(z)$ and consequently the form factors.
As discussed later,
the choice $m=\tau - 3/2$ yields the correct
falloff estimated by perturbative QCD (pQCD)~\cite{Carlson}.
Consequently, the spectrum~(\ref{Leigenvalues}) takes the
following form~\cite{Teramond12a,Gutsche13a,Brodsky15},
\ba
M_{Nn}= 2 \kappa \sqrt{ n + \tau -1}\,.
\label{eqMassN}
\ea
Thus, holography provides an estimate for the spectrum
of the nucleon radial excitations
in terms of a single scale parameter $\kappa$.
Similarly, one obtains
the spectrum of mesons and baryons with different
angular momenta and parity~\cite{Teramond12a,Brodsky15,Teramond15a}.

The $\rho$ meson (mass $m_\rho$)
is a traditional reference for the hadron states.
In LF holography we can write $m_\rho = 2 \kappa$.
The nucleon mass, for example, approximately satisfies $M_N=\sqrt{2}m_\rho$.
If we assume that the nucleon mass is primarily composed
of the leading twist $\tau=3$ contribution, $M_N=2\sqrt{2}\kappa$,
then the phenomenological value of $\kappa$ is fixed at
\be
\label{kappa}
\kappa \ = \ \frac{m_\rho}{2}\sim \Lambda_{\rm QCD}\,.
\ee

The leading twist estimate of the masses of nucleon radial radial excitations
is now given by
$M_{N n}= 2 \kappa \sqrt{n + 2}$, $n=0,1,2...$.
This yields  $ M_R =  M_{N1} = 2 \sqrt{3} \kappa$
for the  Roper mass in leading twist approximation ($\tau=3$).

It is worth mentioning that the LF estimate of the Roper mass,
$M_R \simeq \sqrt{3} m_\rho$, is not so  accurate
as for the nucleon and the $\rho$ meson.
This is consistent with a general expectation that leading twist approximation
is not so accurate for baryons as it is for mesons.
In particular for the Roper, there are indications that the
baryon-meson-meson corrections of the twist order $\tau=7$
are important~\cite{Suzuki10,Gegelia16}.
In the next subsection we collect further comments
on the spectrum and other issues of LF holography.

The spectrum of the hadrons, as predicted
by LF holographic approach has further issues.
A similar approach to mesons yields similar
Regge-behaved spectra as the one given by Eq.~(\ref{eqMassN}).
For the $\rho$ meson
family one gets $m_{\rho n}= 2 \kappa \sqrt{n + 1}$ for
$n=0,1,2,\ldots$~\cite{Grigoryan07b,Branz10a,Karch06}.
Brodsky and Teramond argue, however, that those
twist-2 mass poles should be shifted
to their physical values, suggesting that
$m_{\rho n} = 2 \kappa \sqrt{2n +1}$~\cite{Teramond15a,Brodsky15,Sufian17a}.
These formula gives a very good agreement
with the vector meson masses for
$\kappa \simeq 0.385$ GeV~\cite{Teramond15a,Brodsky15},
as in Eq.~(\ref{kappa}), provided that
$m_\rho  \simeq 770$ MeV. Similar values were used
in Refs.~\cite{Gutsche12a,Gutsche13a}.

\subsection{Comments on interpolating operators in LF holographic approach}

The scaling dimension $\Delta$
is defined by the behavior of the
wave function in the limit $z \to 0$:
$\Psi (x,z) \sim z^\Delta$~\cite{Abidin09,Brodsky08a}.
The holographic approach relates the
5D mass parameter $m$ with the scaling dimension $\Delta$
of the hadron interpolating operator. For example,
for the vector field this relation is
\be
\Delta \ =\ 2+\sqrt{m^2+1}\,.
\ee
For the massless vector field $V_M$ above, this infers $\Delta=3$,
the correct scaling dimension of a current operator~\cite{Brodsky08a}.
For fermions however the relation is
\be
\Delta = 2 + |m|\,.
\ee
In the leading twist this yields $\Delta = 7/2$ for the nucleon/Roper
interpolating operator, in contrast to the scaling dimension
$\Delta_{\cal O}=9/2$ of the operator ${\cal O}=qqq$,
composed of three quark fields.
This discrepancy is even more pronounced if one applies
the identification $m(\tau,L)$ used in Ref.~\cite{Brodsky15}.
Indeed, comparing Eqs.~(\ref{eqFR}) and~(\ref{Leigenvalues})
with Eqs.~(5.34) and~(5.38) of Ref.~\cite{Brodsky15}
one ends up with the relation $m=\nu+1/2$, which implies $\Delta=5/2$.

In QCD the  (anomalous) scaling  dimensions of the interpolating operators
are not well-defined, because of the logarithmic running of
the gauge coupling. In a putative conformal theory,
which would approximate the slow running of the coupling,
the anomalous scaling dimensions would exist and correspond
to a dual gravity theory of the type studied here.
In particular, the massless vector field of a gravity dual
would correspond to $\Delta=3$ conserved symmetry current.
In the meantime we have to accept that the correct choice
of the operator dimension leads to unpleasant effects
like incorrect scaling of the form factors for very large $Q^2$.

One can argue, in principle, that predictions of holographic models
are generally valid in the strongly coupled regime of a theory,
thus the pQCD prediction should not be compared with the outcome
of the holographic analysis.
We believe, however, that this argument does not quite apply
to the scaling of the form factors, since the pQCD prediction
is rather based on the assumption of a particular partonic structure
of the hadron, which is not exactly
the full perturbative quark picture~\cite{Carlson}.

We hope that some more advanced holographic model can resolve
the issues mentioned above. In particular, top-down models,
as more complex ones, naturally share more intricate details with QCD.
A prototypical example would be (the glueball sector of) the
Klebanov-Strassler theory~\cite{Klebanov00,Strassler05}.
Albeit supersymmetric, this theory encodes a logarithmically
running gauge coupling, thus, extracting dimensions of the operators
in this theory requires removing the logarithmic scale dependence
(\emph{e.g.}~\cite{Krasnitz02,Dymarsky08,Gordeli13}).
Moreover, in this theory, the states with the same quantum numbers
tend to mix with each other. Therefore the mass eigenstates
are superpositions of states with different twist.
Mixing has effect of changing the resulting spectrum as well
as shifting the values of the
dimensions~\cite{Berg05,Dymarsky08,Gordeli09,Gordeli13,Gordeli16}.

\section{Nucleon electromagnetic form factors}
\label{secNucleon}

The transition current $J^\mu$ between
two nucleon states (elastic transition)
can be expressed,
omitting the asymptotic spin states (spinors)
and the electric charge $e$, as~\cite{NSTAR,Nucleon}
\ba
J^\mu =
F_{1N} (Q^2)
 \gamma^\mu
+ F_{2N} (Q^2)\frac{i \sigma^{\mu \nu} q_\nu}{2 M_N}\,,
\label{eqNucleonCur}
\ea
where $F_{iN}$ ($i=1,2$)
define, respectively, the Dirac and Pauli form
factors of the proton ($N=p$)
and neutron ($N=n$).
In the LF formalism, the two form factors appear
in a spin-nonflip ($F_{1N}$) and
a spin-flip ($F_{2N}$) transitions~\cite{Brodsky80,Brodsky01}.

The calculation of the overlaps~(\ref{Sint})
in the holographic model with the interaction (\ref{eqVtotal}),
where one replaces $\Psi$ and $\bar\Psi$ by the appropriate initial
and final nucleon state modes, in this case the nucleon modes $\Psi_{N0}$.
The leading twist case ($\tau=3$), the Dirac form factor is determined by
the functions $F_{L,0}^2$  and $F_{R,0}^2$,
and the Pauli form factor is determined by
overlap of the $F_{L,0}$ and $F_{R,0}$ components~\cite{Gutsche12a,Gutsche13a}.
The final expressions for the form factors,
as originally derived in Refs.~\cite{Abidin09,Gutsche12a}, read:
\ba
F_{1N} &=& e_N\frac{a+ 6}{(a+1)(a+2)(a + 3)} +
\nonumber \\
   & & g_V  \delta_N \frac{a}{(a+1 )(a +2)(a +3)} + \nonumber \\
   & & \eta_N\frac{2 a (2a-1)}{(a +1 )(a +2)(a + 3)(a + 4)},
\label{eqF1N} \\
F_{2N} &=&
\eta_N
\frac{M_N}{2 \sqrt{2}\kappa}
\frac{48}{(a +1 )(a +2)(a +3)},
\label{eqF2N}
\ea
where $a= \frac{Q^2}{4 \kappa^2}$ and
$\kappa$ is the holographic scale
discussed in the previous section.
In Eqs.~(\ref{eqF1N})-(\ref{eqF2N}),
$\delta_N= \pm 1$ ($\delta_p=1$, $\delta_n=-1$)
and $\eta_N$ take different values
for proton and neutron.
In Eq.~(\ref{eqF2N}) we can replace $\frac{M_N}{2 \sqrt{2} \kappa}$
by unity, if $2 \sqrt{2} \kappa$ is a good approximation
for the nucleon mass.

\begin{figure*}[t]
\vspace{.5cm}
\centerline{
\mbox{
\includegraphics[width=3.0in]{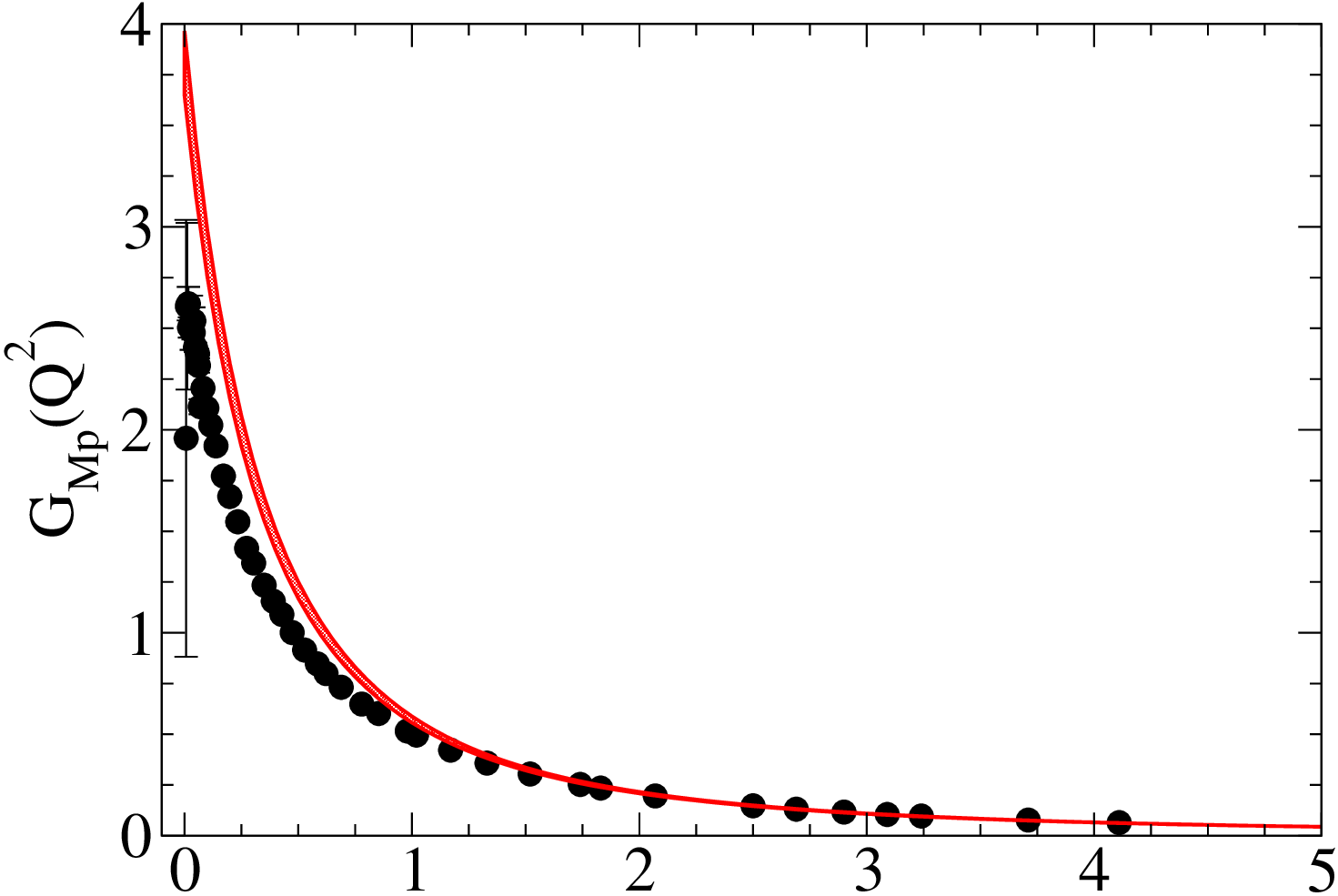}  \hspace{.3cm}
\includegraphics[width=3.0in]{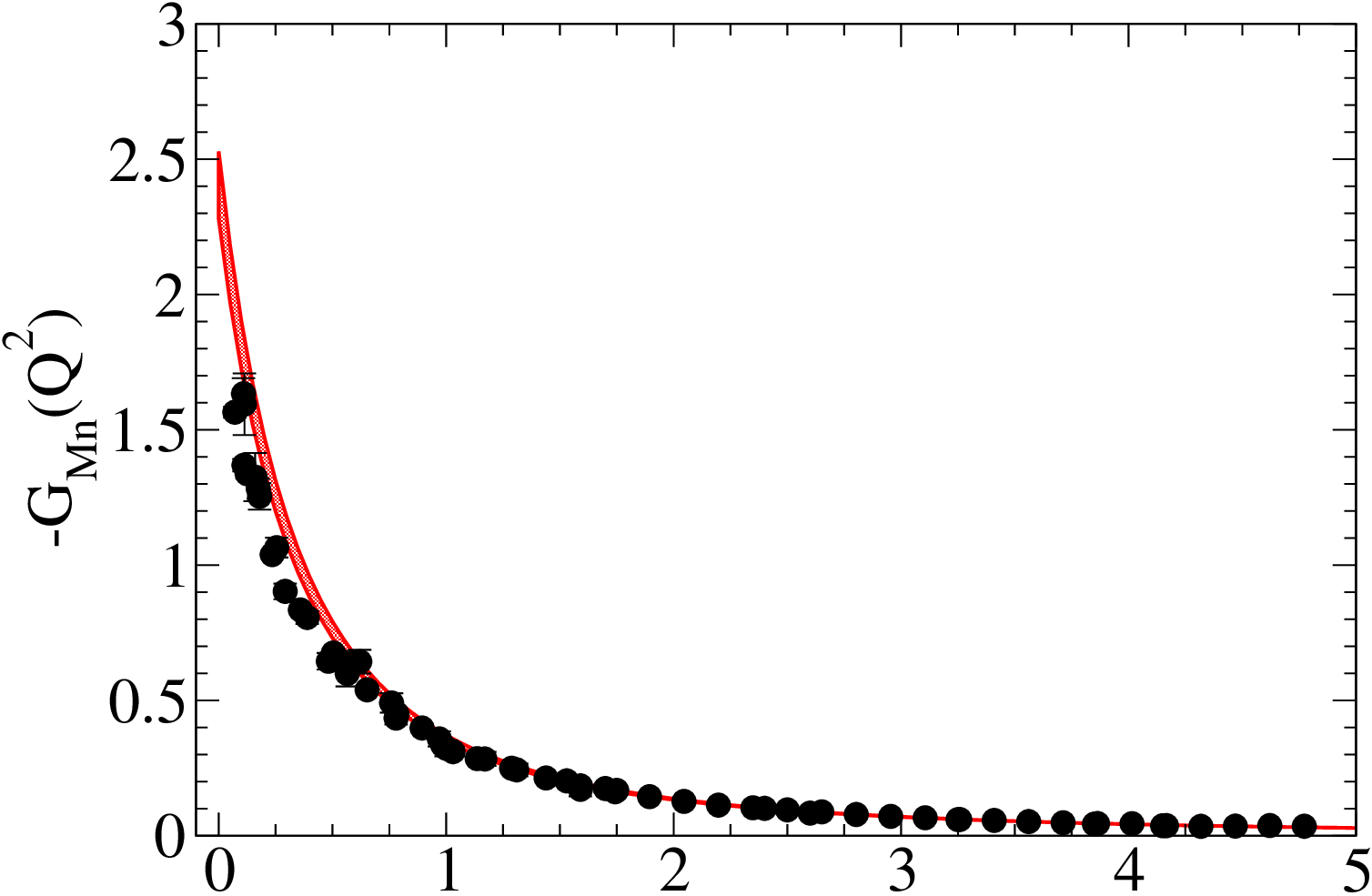}
}}
\vspace{1cm}
\centerline{
\mbox{
\includegraphics[width=3.0in]{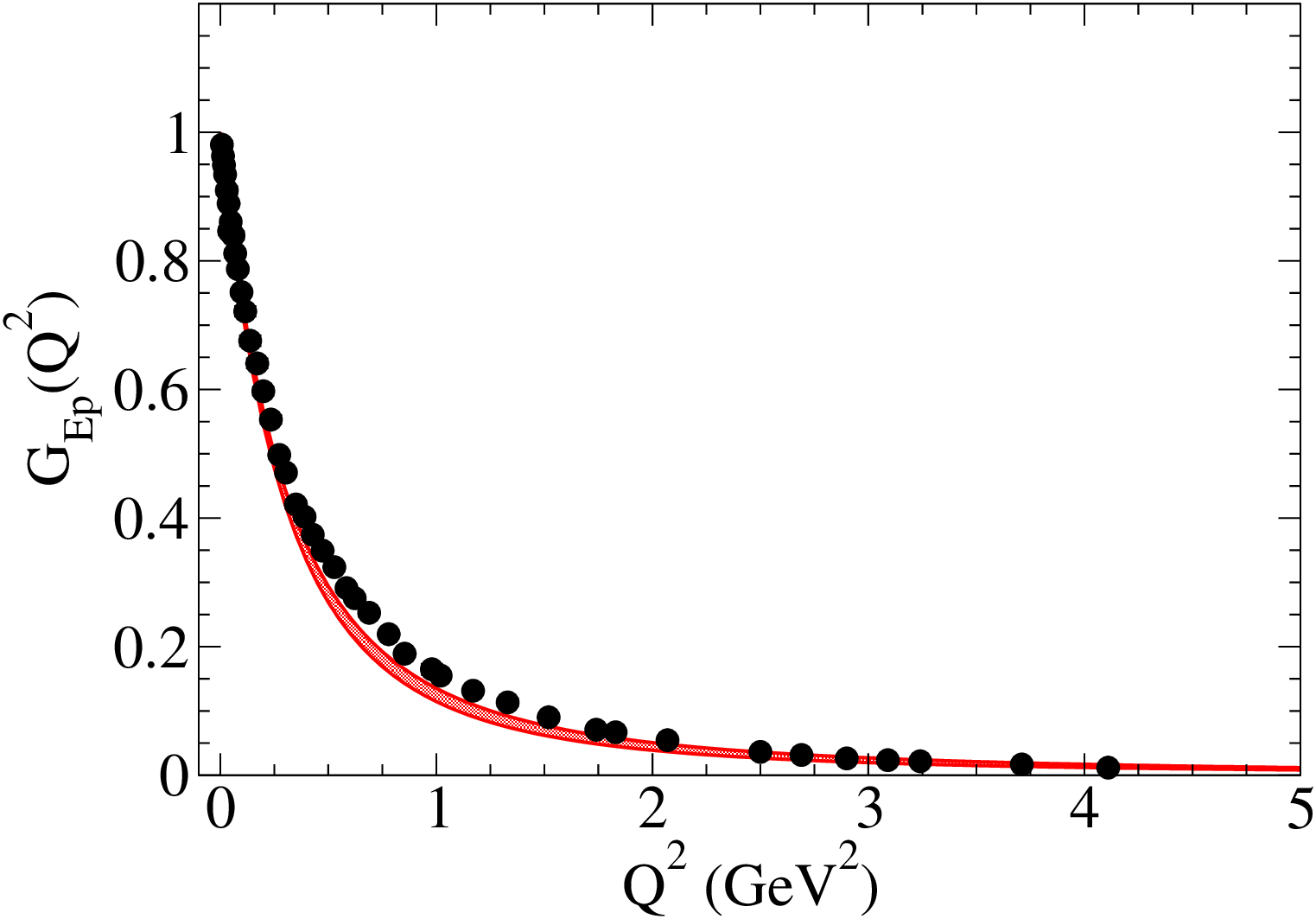}  \hspace{.3cm}
\includegraphics[width=3.0in]{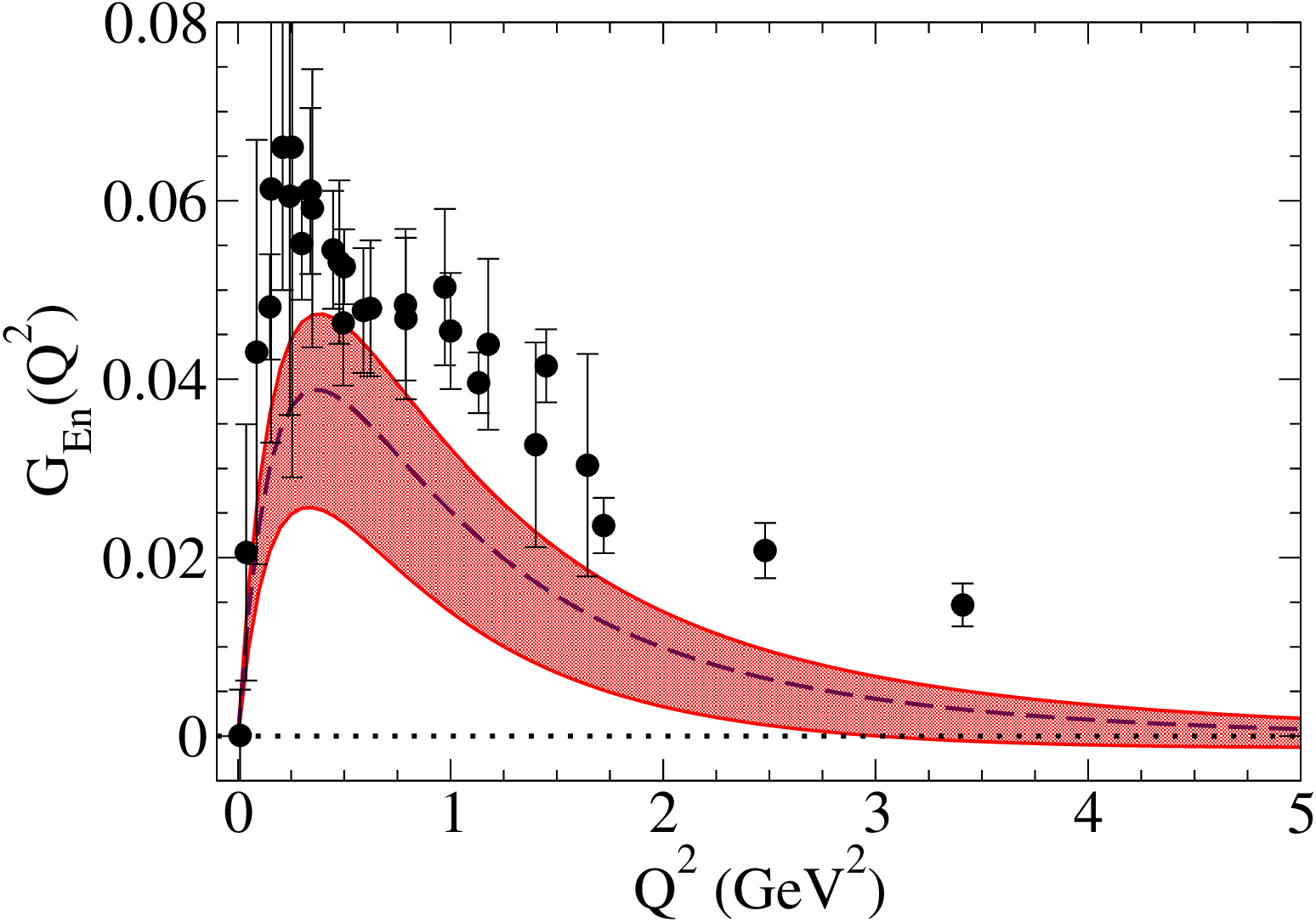}
}}
\caption{\footnotesize{
Proton and neutron electric and magnetic form factors
for the parametrization from Table~\ref{tableTub}.
The red band represents the interval between
parametrization with $Q_m^2=1.5$ GeV$^2$ and $2.5$ GeV$^2$.
In the case of $G_{En}$, the dashed-line 
represents the parametrization with $Q^2_m=2.0$ GeV$^2$.
Data from Refs.~\cite{ProtonData,RefsGEn,RefsGMn}.
}}
\label{figGEGM}
\end{figure*}

In an exact $SU(2)$-flavor model one has $\eta_n= -\eta_p$.
These parameters
can be determined by the proton and neutron
bare anomalous magnetic moment $\kappa_p^b$ and $\kappa_n^b$
(we use the upper index $b$ to indicate
the bare values).
In a model with no meson cloud one can write
$F_{2p}(0) \equiv  \kappa_p= 8 \eta_p$
and $F_{2n}(0) \equiv  \kappa_n = 8 \eta_n$,
assuming $M_N = 2 \sqrt{2} \kappa$.

The holographic results~(\ref{eqF1N})-(\ref{eqF2N}) yield
the correct perturbative QCD behavior for the
nucleon electromagnetic form factors:
$F_{1N} \propto 1/Q^4$ and $F_{2N} \propto 1/Q^6$,
as consequence of $m= 3/2$.
Another observation about the current LF model and results~(\ref{eqF1N})-(\ref{eqF2N})
is that contrary to a common viewpoint on vector meson dominance in top-down holography,
the number of poles contributing to the form factors is finite.
This would not be the case, however, if a generic coupling
$\hat{\mathcal{V}}$ is used in the overlap integral~(\ref{Sint}).
For a more detailed discussion
see Refs.~\cite{Abidin09,Gutsche12a,Gutsche13a,inpreparation}.

As previously discussed the couplings $\eta_N$
and $g_V$ are included phenomenologically.
As can be seen from Eqs.~(\ref{eqF1N})-(\ref{eqF2N}),
the minimal coupling
only produces the Dirac form factor.
The non-minimal couplings generate a
non-zero $F_{2N}$
and adds an extra contribution to the Dirac form factor
(proportional to $\eta_N$)~\cite{Chakrabarti14a}.
Also the minimal-type coupling $g_V$ give
a contribution for the Dirac form factor.
Relations of the type~(\ref{eqF1N})-(\ref{eqF2N})
were derived for the first time in Ref.~\cite{Abidin09}
for the case $g_V=0$.

To determine the values of $g_V$
and $\eta_N$ we performed
a fit to the data $Q^2 \ge Q^2_m$
where $Q_m^2$ is the threshold
of the data included in the fit.
We considered the cases $Q^2_m=1.5$, 2.0 and 2.5 GeV$^2$,
since the meson cloud effects are in principle
suppressed for large $Q^2$,
but we do not know a priory the threshold of the
suppression.
We use the data from Refs.~\cite{ProtonData,RefsGEn,RefsGMn}.
Check Refs.~\cite{Octet,Medium} for a detailed
description of the database.

From the fits, we conclude that the results for
the neutron electric form factor $G_{En}$
are very sensitive to the value of $Q_m^2$.
If the low $Q^2$ is included in the fit ($Q^2_m < $ 1.0 GeV$^2$)
the best fit favors a negative $G_{En}$ for low $Q^2$,
in conflict with the measured data.
This result is very pertinent,
because it is known that meson cloud effects,
in particular the pion cloud, is very
important for the description of the $G_{En}$ at low $Q^2$.
Since the pion cloud contributions are
not included in the  parametrization
of Eqs.~(\ref{eqF1N})-(\ref{eqF2N}),
it is not surprising to see that
the fit with $Q_m^2=1.5$ GeV$^2$ fails
to describe the $G_{En}$ data.
It can be surprising, however, to note
that the best description of the nucleon data
occurs for $Q^2 > 1.5$ GeV$^2$ in the fits
with $Q_m^2=2.0$ and 2.5 GeV$^2$,
when only a few data points for $G_{En}$ are considered.
One then concludes  that
the shape of $G_{En}$ is determined by
the form factors $G_{Ep}$, $G_{Mp}$ and $G_{Mn}$.
This result shows the consistence of the fitting procedure.

The parameters obtained for the fits with
$Q^2_m=1.5$, 2.0 and 2.5 GeV$^2$ are presented in
Table~\ref{tableTub}.
We can conclude that the
parameters are sensitive to the
data included in the fit.
In particular, $g_V$ depends strongly
of the threshold $Q_m^2$.

\begin{table}[t]
\begin{center}
\begin{tabular}{c c c c}
\hline
\hline
$Q_m^2$(GeV$^2$) & $g_V$ & $\eta_p$  & $\eta_n$ \\
\hline
1.5  &  1.571\spQ &  0.378\spQ &  $-$0.326 \\
2.0  &  1.370\spQ &  0.410\spQ & $-$0.345 \\
2.5  &  1.275\spQ &  0.424\spQ & $-$0.361 \\
\hline
\hline
\end{tabular}
\end{center}
\caption{Parameters $g_V$, $\eta_p$ and $\eta_n$
obtained by the fit of the nucleon form factor data
with $Q^2 \ge Q_m^2$.}
\label{tableTub}
\end{table}

The results obtained for the nucleon
electric ($G_{EN} = F_{1N} - \sfrac{Q^2}{4M_N^2} F_{2N}$)
and magnetic ($G_{M\!N} = F_{1N} +  F_{2N})$
form factors using  the parameters from Table~\ref{tableTub}
are presented in Fig.~\ref{figGEGM}.
In the figure, we include a band
to represent the interval between the models
with $Q_m^2=1.5$ GeV$^2$ and $2.5$ GeV$^2$.
In the calculations we used $\kappa=0.385$ GeV,
in order to have a good description
of the $\rho$ mesons and nucleon masses.

In general the estimates from different parametrizations
are very close, except,
for the neutron electric form factor.
The function $G_{En}$ is in fact very sensitive
to the parameters $g_V$, $\eta_p$ and $\eta_n$.
In the graph of $G_{En}$ the lower limit
corresponds to $Q_m^2=1.5$ GeV$^2$ and the
upper limit to the case $Q_m^2=2.5$ GeV$^2$.
The intermediate case ($Q_m^2=2.0$ GeV$^2$)
is included in order to emphasize the strong dependence
of $G_{En}$ with the threshold $Q_m^2$ used in the fit,
and it is represented by the dashed-line.

Concerning the remaining form factors,
one can notice an overestimation
of the low $Q^2$ data ($Q^2 < 1$ GeV$^2$) for $G_{Mp}$ and $-G_{Mn}$.
These results may be interpreted as the manifestation
of the pion cloud effects,
not included in the formalism associated with
Eqs.~(\ref{eqF1N})-(\ref{eqF2N}).

In the past~\cite{Gutsche12a,Gutsche13a,Chakrabarti14a}
the parameters $\eta_N$ and $g_V$ have been fixed
by the static properties of the nucleon,
such as the anomalous magnetic moments
and the nucleon electric charge radius.
Since those observables depend on the meson cloud, we choose
in this work to fix the parameters
in a region where the meson cloud is
significantly reduced,
in order to obtain a more accurate estimate
of the bare parameters.

In the recent work the Brodsky-Teramond model
for the nucleon~\cite{Sufian17a} was improved
with the inclusion of the explicit pion cloud contributions
($q \bar q$ states) with a few adjustable parameters.
It was shown that, indeed, the inclusion
of the pion cloud contribution ($\tau=5$) is
fundamental for an accurate description of $G_{En}$.

In order to understand the role
of the valence quark degrees of freedom
we restrict the present study to the
the leading twist contribution ($\tau=3$).
Once fixed the parameters of the model (bare parameters)
one can use those parameters to calculate the
valence quark contributions for the
$\gamma^\ast N \to N(1440)$ transition form factors.

\section{$\gamma^\ast N \to N(1440)$ form factors}
\label{secRoper}

We consider now the $\gamma^\ast N \to N(1440)$
transition form factors.
Again omitting the spinors of the nucleon and the Roper
(and the electric charge $e$), the transition current $J^\mu$
can be expressed as~\cite{Roper}
\ba
J^\mu =
F_{1N}^\ast(Q^2)\left(
 \gamma^\mu  - \frac{{\not \! q} q^\mu}{q^2} \right)
+ F_{2N}^\ast (Q^2)\frac{i \sigma^{\mu \nu} q_\nu}{M_N + M_R},
\nonumber \\
\label{eqRoperCur}
\ea
where $F_{1N}^\ast$ and $F_{2N}^\ast$
 ($N=p,n$) are
the Dirac and Pauli transition form factors, respectively.

Explicit analytical calculation, using the holographic
 wave functions~(\ref{PosChirMod}) and~(\ref{NegChirMod}) for
the nucleon and the Roper,
combined with the interaction (\ref{eqVtotal}), shows
that the Dirac and Pauli form factors associated
with the $\gamma^\ast N \to N(1440)$
in leading twist~\cite{Gutsche13a} are
\ba
F_{1N}^\ast &=& \delta_N
\frac{a (\sqrt{2} a + c_1) }{(a+1)(a+2)(a+ 3)(a + 4)} +
\nonumber \\
   & & g_V  \delta_N \frac{a (\sqrt{2} a + c_2) }{(a+1)(a+2)(a+ 3)(a + 4)} +
\nonumber  \\
   & & \eta_N\frac{2 a ( 2 \sqrt{2} a^2  -c_3 a + c_4)}{(a+1)
(a+ 2)(a + 3)(a+ 4)(a+5)},  \label{eqF1R} \\
F_{2N}^\ast &=&
\eta_N
\left(\frac{M_R + M_N}{M_R}\right)^2
\frac{M_R}{2\sqrt{3}\kappa} \nonumber \\
& & \times
\frac{6 \sqrt{3} (c_5 a -4)}{(a + 1)(a + 2)(a +3)(a+4)},
\label{eqF2R}
\ea
where $c_1=4 \sqrt{2} + 3 \sqrt{3}$,
$c_2=4 \sqrt{2} - 3 \sqrt{3}$,
$c_3= 9 (\sqrt{3} -\sqrt{2})$,
$c_4= 3 \sqrt{3}- 5 \sqrt{2}$
and $c_5=2 + \sqrt{6}$.

In Eq.~(\ref{eqF2R}) we can replace
$\sfrac{M_R}{2\sqrt{3}\kappa}$ by unity if
$2\sqrt{3}\kappa$ is a good approximation to the
Roper mass.
In the original form, Eqs.~(\ref{eqF1R}) and~(\ref{eqF2R})
were obtained in Ref.~\cite{Gutsche13a}.
In comparison with that work
we included an extra factor $\sfrac{M_R + M_N}{M_R}$
in order to be consistent with the
more usual definition
of the transition form factors~(\ref{eqRoperCur}).

\begin{figure*}[t]
\vspace{.5cm}
\centerline{
\mbox{
\includegraphics[width=3.2in]{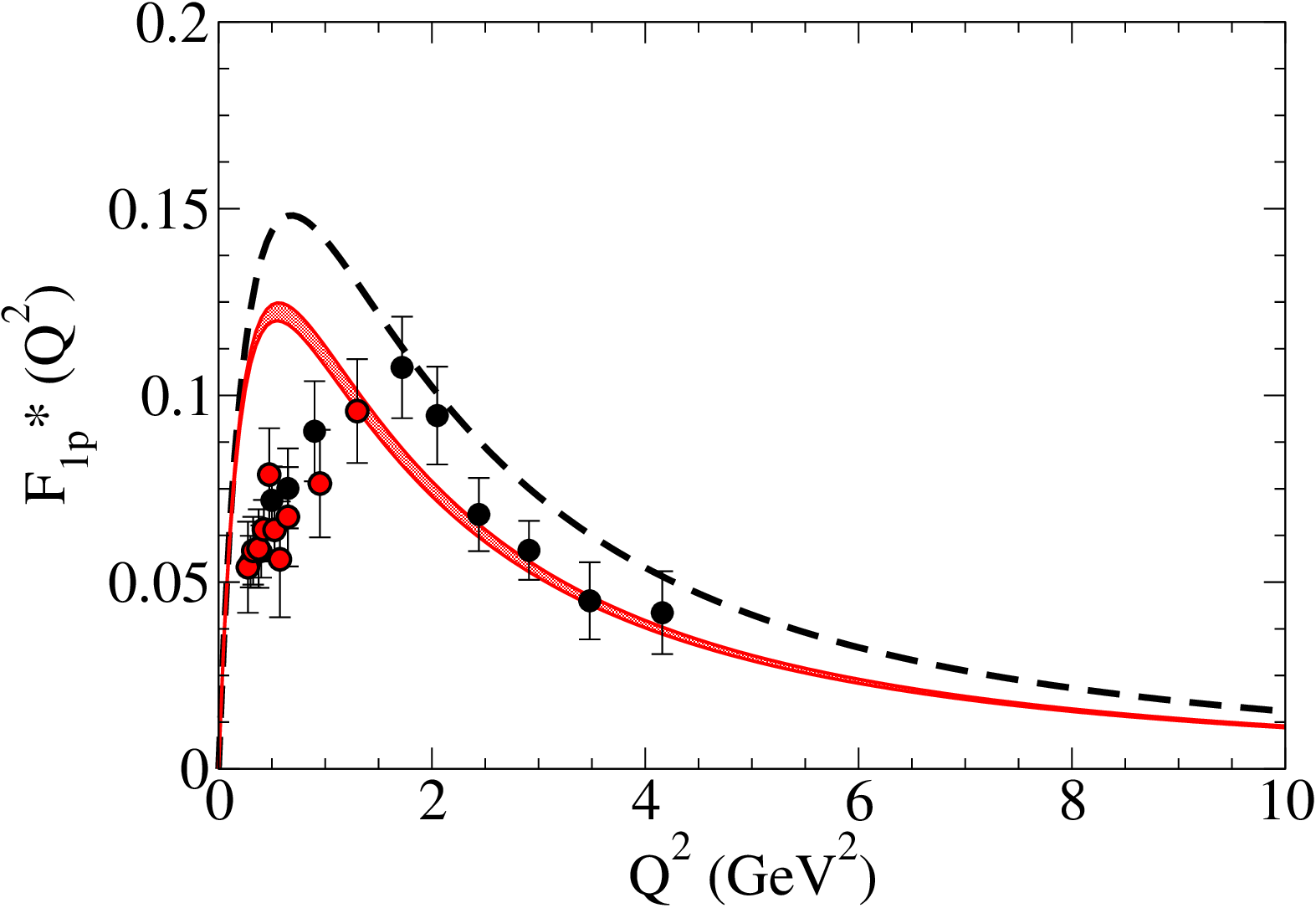}  \hspace{.1cm}
\includegraphics[width=3.2in]{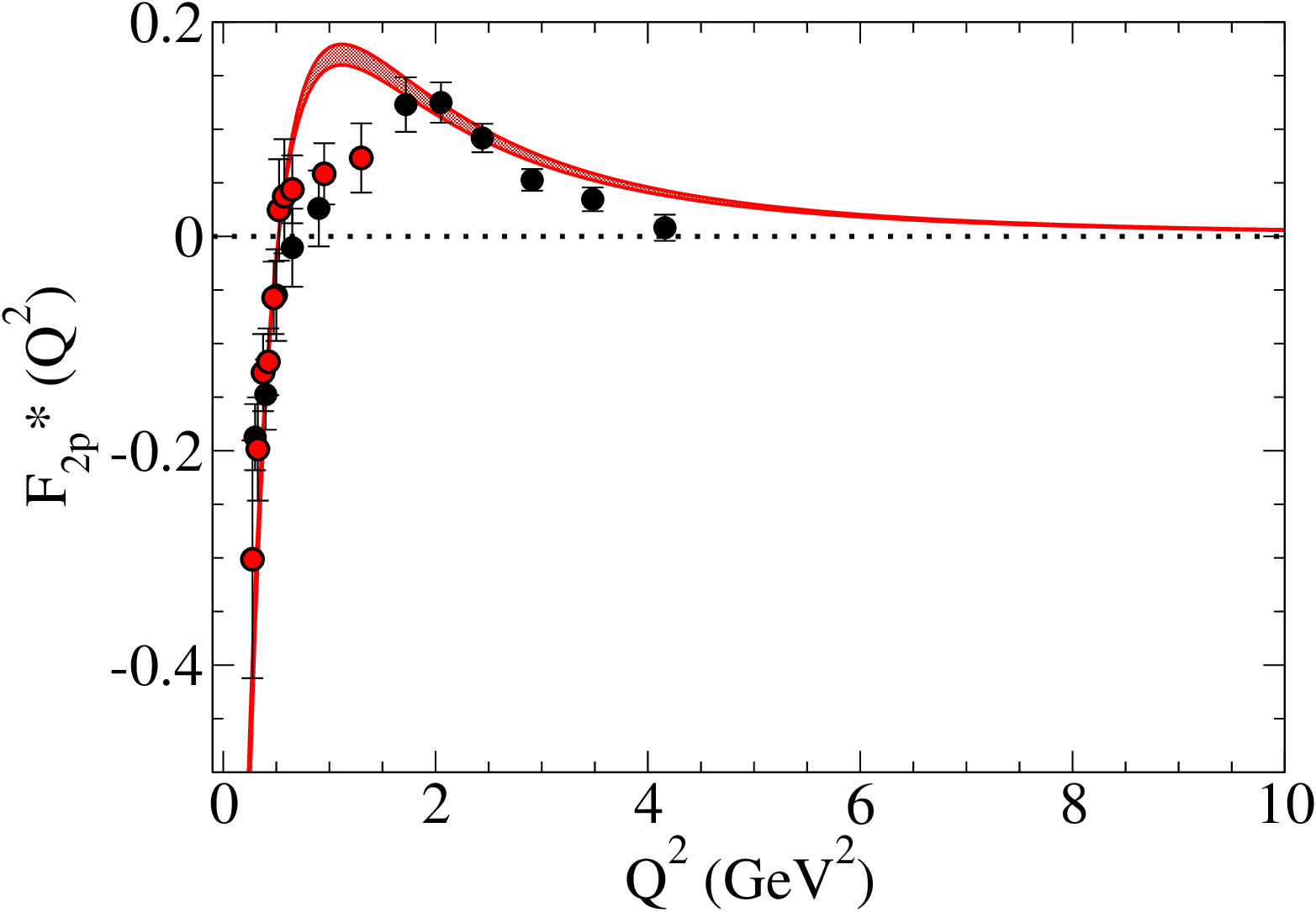}
}}
\caption{\footnotesize{
$\gamma^\ast N \to N(1440)$ transition
form factors $F_{1p}^\ast$ and  $F_{2p}^\ast$.
The red band represents the interval between
the models with  $Q_m^2=1.5$ GeV$^2$ and $2.5$ GeV$^2$.
The dashed-line is the result
from the Light-Front holographic
model from Ref.~\cite{Teramond11a}.
Black dots from Ref.~\cite{CLAS1};
red dots from Ref.~\cite{CLAS2}.}}
\label{figRoper}
\end{figure*}

The results for the nucleon to Roper transition
form factors associated with the parameters
$g_V$, $\eta_p$ and $\eta_n$ discussed previously
(Table~\ref{tableTub}), are presented in Fig.~\ref{figRoper},
for the proton target ($N=p$).
As for the nucleon we use $\kappa =0.385$ GeV.
The data presented
here are those from CLAS/Jefferson Lab
for single pion production~\cite{CLAS1}
and double pion production~\cite{CLAS2}, and
it is collected in the database~\cite{MokeevDatabase}.
We do not discuss the results for the neutron target ($N=n$)
because they are restricted to the photon point.

In Fig.~\ref{figRoper}, we can see that,
contrary to the nucleon case,
the results are almost insensitive
to the variation of parameters, therefore we consider
only the upper and lower cases ($Q_m^2=1.5$ and 2.5 GeV$^2$),
delimited by the band.
In the figure one can see
that the  holographic approach  based on
Eqs.~(\ref{eqF1R})-(\ref{eqF2R})
in the leading twist approximation
gives a very good description of the large $Q^2$ data
($Q^2 > 1.8$ GeV$^2$) in  the range
$g_V=1.28$,...,1.57, $\eta_p= 0.38$,...,0.42
and $\eta_n =-(0.36$,...,$0.32)$.

In the above holographic approach
one obtains a clear estimate
of the valence quark contributions
based on framework that includes only one
pre-defined parameter: the mass scale $\kappa$.
Since the coefficients $g_V$, $\eta_p$, $\eta_n$
are determined by the nucleon form factors,
one can consider this estimate of the
valence quark contributions for
the nucleon to Roper form factors as a
parameter-free prediction.

The present results for $F_{1p}^\ast$ and $F_{2p}^\ast$
are consistent with estimates of the form factors based
on different frameworks, such as the results of
Refs.~\cite{Aznauryan07,Roper,Roper2,Segovia15,Aznauryan12b}.
In particular, the present model
and the results of the covariant quark model
from Refs.~\cite{Nucleon,Roper,Roper2}
are very close for $Q^2 > 5$ GeV$^2$.

The holographic calculation
for the nucleon to Roper transition form factors
given by Eqs.~(\ref{eqF1R})-(\ref{eqF2R})
were derived originally in Ref.~\cite{Gutsche13a}.
In that work, higher Fock states (twist 4 and 5 contributions)
were also taken into account.
However the effective contribution of the higher Fock
states was determined by the equation for the Roper mass,
and not by the physics associated with the form factors
(dominance of the leading twist contributions for large $Q^2$).
In addition, the overlap between the nucleon and Roper states
was determined in Ref.~\cite{Gutsche13a} using coefficients
adjusted to the form factor data.

In the present work we rather prefer to concentrate on deriving
a clean estimate of the valence quark contribution
using an alternative method of fixing the parameters of the model,
instead of trying to estimate the quark-antiquark
and gluon contributions using LF holography.
Since we do not consider higher twist contributions,
the overlap between the states is just the result
of the overlap between the valence quark states of both baryons
with no unknown coefficients.

Although we conclude that
the leading twist calculation is
a good approximation for the form factors,
one still notes that it may not be sufficient
for a satisfactory estimate for the mass of the Roper.
Different estimates of the mass suggest,
in fact, that higher Fock states
are crucial for the explanation of
the experimental value~\cite{Suzuki10,Capstick00,Gegelia16}.

In Fig.~\ref{figRoper},
we also present the first estimate of the
the Dirac form factor, $F_{1p}^\ast$, performed by
Teramond and Brodsky~\cite{Teramond11a}.
In their formulation, the Dirac form factor
is expressed by an analytic function dependent on
the $\rho$ meson masses~\cite{Brodsky15,Teramond11a,Teramond12a}.
The closeness between results
is a consequence of the use of the holographic masses
instead of the physical masses.
We note at this point, that,
the relations (\ref{eqF1R})-(\ref{eqF2R})
are also analytic parametrization
of the Roper form factors, since the
form factors are expressed in terms of functions
of $a= \frac{Q^2}{m_\rho^2}$,
using $m_\rho^2 = 4 \kappa^2$.
We leave however the comparison
between analytic parametrization
of the Roper form factors to
a separated work~\cite{newPaper}.

In the graph for $F_{1p}^\ast$
one can notice a deviation between
our estimate and the data for $Q^2 < 1$ GeV$^2$.
This result can be interpreted
as the consequence of the meson cloud
contributions not included
in our leading twist analysis.
Similar results were obtained in
independent works~\cite{Aznauryan07,Roper,Segovia15}.
Sizable contributions
of the meson cloud effects were found in
Refs.~\cite{Golli09,Obukhovsky11}.

As for the results for $F_{2p}^\ast$,
one can note that the parametrization
based on the LF holography
gives negative values for small $Q^2$,
and are close to the data.
In particular we estimate the
change of sign in $F_{2p}^\ast$
for $Q^2 \approx 0.5$ GeV$^2$.
The results for  $F_{2p}^\ast$ suggest that contrarily to
the form factor $F_{1p}^\ast$, where  the meson cloud are sizable,
for  $F_{2p}^\ast$ the meson cloud contribution
are very small at low $Q^2$.
To the best of our knowledge, it is the first time
that this fact was observed in the context of a quark model.
There is, nevertheless, a discrepancy
between the model and the data in the region between 0.9 and 1.5 GeV$^2$.

Concerning the $F_{2p}^\ast$ data,
one can note that in the region 0.9--1.5 GeV$^2$ there
are only two datapoints associated
the two pion electroduction CLAS data~\cite{CLAS2}.
It is important to check
if the new data associated with the one pion production
confirms the trend from Ref.~\cite{CLAS2}.
At the moment one can conclude that
the holographic model gives a very good description
of the $F_{2p}^\ast$ data except for three data points
in the range 0.9--1.5 GeV$^2$
(underestimation of the data in about 60\%).

\subsection*{Discussion}

Summarizing the results presented here
for the nucleon and Roper form factors,
one can conclude
that the leading twist approximation provides
a very good estimate for both
nucleon and Roper form factors.
This result is consistent with the
LF formalism, because
in the Drell-Yan-West frame the contributions
from higher Fock states are small~\cite{Brodsky08a,Perdrisat07,Brodsky15}.
It is expected however that the meson cloud effects
provide significant contributions
to the $\gamma^\ast N \to N^\ast$ transition form factors
for some nucleon excitations $N^\ast$
for small $Q^2$~\cite{NSTAR,S11,SQTM,SemiRel,NDelta-Lattice,Delta1600,D13,DeltaQuad,OctetDecuplet,DeltaTL}.

In the present work, the meson cloud effect
is observed in particular for the
form factors of the Roper: $F_{1p}^\ast$, and
in the region 0.9--1.5 GeV$^2$ for  $F_{2p}^\ast$.
Those meson cloud corrections come
from higher twist contributions to
the Light-Front wave functions.
For higher mass resonances it is expected
that higher twist corrections also give
 important contributions to the
transition form factors at low $Q^2$.
In that case LF holography
can be used to estimate mainly
the contribution of the quark core.
Although the estimate for low $Q^2$ may
appear rough, since the calculation
is based on massless quarks
(and the physical quarks have mass),
for large $Q^2$ the estimate is expected to be 
accurate due to the dominance of the valence quark degrees of freedom.

In the future one can use holography
to estimate transition form factors
in the leading twist approximation for other $N^\ast$ states.
Those estimates are expected to be accurate
at large $Q^2$, since they are based on
valence quark degrees of freedom.
At low $Q^2$ LF holography may fail in leading twist,
since the framework provides only the valence quark contributions
to the form factors.
That information can be, however,
very useful to understand the role of the meson cloud contributions.
On one hand, we can use the comparison
with the data to estimate the effect of
the meson cloud contribution.
On the other hand, estimates of the bare core can also
be used as input to dynamical coupled-channel reaction models
in the parametrization of baryon bare core~\cite{Suzuki10,EBAC,Burkert04}.

It is worth mentioning that some authors interpret
the Roper as a dynamically-generated
baryon-meson resonance,
without an explicit reference to three-quark systems,
except for the nucleon and
the $\Delta(1232)$~\cite{Krehl00,Doring09,Liu16a,Lang17}.
As far as we know, there are no calculations
of $\gamma^\ast N \to N(1440)$ transition form factors
for intermediate and large $Q^2$,
based on dynamically-generated resonance models for the Roper.
Future lattice QCD simulations can help
to understand the role of the baryon-meson
contribution for the transition form factors at
low $Q^2$~\cite{Liu16a,Lang17}.

\section{Outlook and conclusions}
\label{secConclusions}

In the present work we apply the LF holography
in the soft-wall approximation
to the study of the electromagnetic structure
of the nucleon and nucleon excitations.
More specifically we study the nucleon and
$\gamma^\ast N \to N(1440)$ transition form
factors in the leading twist approximation.

Since in the leading twist approximation the
transition form factors are determined by
the valence quark degrees of freedom,
in the present approach we estimate
the contribution of valence quarks to
the electromagnetic form factors.
The Light-Front wave functions
are determined by an appropriate
choice of boundary conditions in the 5D space
and by the expected pQCD falloff for the
Dirac and Pauli form factors (bottom-up approach to QCD).

The Light-Front holography in the
soft-wall version was used previously
in the study of the nucleon elastic form factors,
in leading twist and higher orders.
However, since in those works the couplings are
adjusted to the dressed couplings
(empirical anomalous magnetic moments)
they may describe well the low $Q^2$ data,
but fail in the description of the large-$Q^2$ data.
In the present work we fix
the free parameters of the model associated
with the bare couplings using the large $Q^2$ data for the nucleon,
where the effect of the meson cloud contributions
is significantly reduced.

Our expressions for the electromagnetic form factors
depend only of the holographic mass scale $\kappa$
and of three bare couplings: $g_V$, $\eta_p$ and $\eta_n$.
Once determined the bare couplings by the nucleon data,
the model is used to predict the
$\gamma^\ast N \to N(1440)$ transition form factors.

Our results for the $\gamma^\ast N \to N(1440)$ transition form factors
compare well with the empirical data.
For the Dirac form factor,
the deviation observed for $Q^2 < 1.0$ GeV$^2$
is compatible with the interpretation
that meson cloud contributions are important in that region.
As for the Pauli form factor,
our estimate is very close to the data,
both at low and at large $Q^2$,
except for 3 datapoints in the range 0.9--1.5 GeV$^2$.
The result at low $Q^2$, suggests
that the meson cloud contributions for $F_{2p}^\ast$ are small.
As far as we know this is the first time
that this effect is observed.

The method used in the present work for the Roper
can in the future be extended for higher mass nucleon excitations.
The use of the Light-Front holography
in leading twist provides then a natural method
to estimate the valence quark contributions
for the transition form factors.
The effect of the meson cloud can then be estimated
from the comparison with the experimental data.

Theoretical estimates of the bare core contributions
are very important for the study of the
baryon-meson reactions
and nucleon electroproduction reactions.
The results from Light-Front holography
may be used  as input to dynamical coupled-channel reaction models
in the theoretical study of those reactions.


\begin{acknowledgments}
G.~R.~thanks Valery Lyubovitskij for useful clarifications about
the calculation of the transition form factors.
This work was supported by the
Universidade Federal do
Rio Grande do Norte/Minist\'erio da Educa\c{c}\~ao (UFRN/MEC).
The work of D.~M. was also partially supported by the RFBR grant 16-01-00291.
\end{acknowledgments}



\end{document}